\documentstyle{article}
\begin{document}
[\begin{center}
\begin{huge}
\baselineskip=10mm plus 1mm minus 1mm
{\bf When did vacuum energy of the Universe become cosmological constant?}\\
Vladimir Burdyuzha\\
\end {huge}
\begin{Large}
Astro-Space Center, Lebedev Physical Institute,\\
Russian Academy of Sciences, Profsoyuznaya 84/32,\\ 117997 Moscow, Russian Federation
\end{Large}
\end{center}]

    A quark-gluon phase transition in the Universe is researched 
  after which vacuum (dark)\\energy has hardened and become cosmological constant.
  Before this a vacuum component of the Universe was changing by jumps during phase
  transitions since vacuum condensates of quantum fields carried a negative contribution
  in its positive density energy. This quintessence period of the
  Universe life took place during the first parts of a second when our Universe
  was losing high symmetry. Using Zel'dovich's formula the modern value of vacuum
  energy is also calculated. It is shown that a quantum chromodynamical vacuum which
  is characterized by pseudogoldstone bosons existed definitely when temperature
  of the Universe was T $\sim 150$ MeV. Therefore there is a large probability  that
  dark energy is vacuum energy.

\twocolumn

     A new experimental fact is dark energy was already boosting the expansion
 of the Universe as long as nine billions years ago. Investigators using NASA's
 Hubble Space Telescope announced this result based on an analysis of the 24
 most distant supernovae during the last two years [1]. In this letter I would
 like to add this experimental fact by a theoretical one that is to find the
 moment when this vacuum energy became the cosmological constant.

     Probably in the time of the Universe birth the temperature was near $\sim 10^{32}$ K
 ($10^{19}$ GeV) and symmetry was very high. During cooling the Universe was losing
 symmetry by phase transitions. Of course, other points of view on an initial
 evolution of the Universe exist. Among them one of the most successful models is a slowly
 varying scalar field, called quintessence [2-3]. L.Randall [4] suggested that the
 Universe evolved naturally in the available vacuum for its. S.Hawking and T.Hertog [5]
 suggested that in the first instants the Universe was in the state of superposition with all
 possible universes. But in any case our Universe has got cooling owing to expansion and
 phase transitions were probably reality (although up to present phase transitions are not
 the standard point of view).

     The simplest chain of relativistic phase transitions (RPT) which was at
 first written in our paper [6] might be realized in very early Universe but
 nobody knows an exact chain of these transitions. The low energy part of
 this chain which must be realized in any case is:

\vspace{0.1cm}

$$
\Rightarrow D_{4}\times SU(3)\times U(1)\Rightarrow D_{4}\times U(1)\eqno(1)
$$
$$
\hspace{0.0cm}...100 GeV.......................150 MeV
$$
\vspace{0.1cm}

 here: $D_{4}, SU(3), U(1)$ are groups of symmetry($D_{4}$ is a group of diffeomorphisms
 relating to gravitation). Let me to research a quark-gluon phase transition
 ($T \sim 150$ Mev) more detail. This phase transition was probably the last one after
 which a vacuum component of the Universe has hardened and this component has
 become cosmological constant.
      Of course, for cooling of cosmological plasma during RPT vacuum condensates
 of quantum fields with negative energy density are produced. In the standard
 model two objects with the same status but different physical properties take
 place: a Higgs condensate and a nonperturbative vacuum condensate. These
 condensates have asymptotic equation of state $p=-\epsilon$. Thus, RPT series
 (a part of which is written in equation (1)) have been accompanied by the
 generation of negative contributions in the initial positive $\Lambda-term$.
 Note that $\Lambda\approx 0$ is best suited to our Universe. A universe with large
 negative $\Lambda$ never become macroscopic. If $\Lambda$ is large and positive
 then production of complex nuclear, chemical and biological structures is impossible.
 Our Universe with hierarchy of cosmological structures can exist only when $\Lambda\approx 0$.

     The present vacuum in the Universe is the vacuum condensate of the last RPT (1)
 which took place when temperature of the Universe was near 150 MeV that is it was
 a quantum chromodynamical (QCD) phase transition [7]. As it is known the chiral
 QCD symmetry $SU(3)_{L}\times SU(3)_{R}$ is not exact and pseudogoldstone bosons
 are the physical realization of this symmetry breaking. The spontaneous breaking
 of this symmetry leads to appearance an octet of pseudoscalar Goldstone states in
 spectrum of particles. For temperature of chiral symmetry breaking $(T_{c}\sim 150 MeV)$
 the main contribution in periodic collective motions of a nonperturbative vacuum
 condensate contributed $\pi$ mesons as the lightest particles of this octet. In this
 process $\pi$ mesons are excitations of the ground state and they characterize this
 ground state definitely that is they characterize QCD vacuum.

     40 years ago Ya.Zel'dovich [8]attempted to calculate a nonzero vacuum energy of
 our Universe in terms of quantum fluctuations of particles as a high order effect.
 He inserted the mass of proton or electron in found them formula but the result was
 not satisfactory. The situation is changed when in his formula to insert the mean
 mass of $\pi$-mesons $m_{\pi}=(2m_{\pi^{\pm}} + m_{\pi^{0}})/3$ = 138,04 MeV

\vspace{0.1cm}
$$
\Lambda=8\pi G^{2} m_{\pi}^{6} h^{-4}\eqno(2)
$$
\vspace{0.1cm}

 here: h is Planck constant; G is gravitational constant (N.Kardashev [9] suggested
 also to use a formula of Zel'dovich for calculation of $\Lambda-term$ modern value). Then

\vspace{0.1cm}
$$
\Omega_{\Lambda} = \rho_{\Lambda}/\rho_{cr} \equiv \Lambda c^{2}/3H_{0}^{2}\eqno(3)
$$
\vspace{0.1cm}

 can be calculated. If $H_{0}=72.5$ (km/sec)/Mpc then $\Omega_\Lambda\simeq 0.7$
 Therefore, it is very probably that in the present epoch the vacuum energy of the
 Universe is the vacuum condensate of the last RPT. In the paper [10] we have already
 calculated $\Omega_{\Lambda}$ for different $H_{0}$ but these calculations and ideas
 are more actually now in the connection with the outstanding experimental fact [1].

      Thus, vacuum energy in first parts of a second of the Universe life was similar
 to quintessence (it changed by jumps during phase transitions when the Universe was
 losing symmetry during cool down from temperature $10^{19} GeV$ till $0.15 GeV$). Of course,
 we do not know an exact chain of Universe phase transitions except last two. For
 temperature $\sim 0.15 GeV$ this vacuum has hardened when the quark-gluon vacuum condensate
 has produced that is it has become cosmological constant.
 Very probably that dark energy is the vacuum component of the Universe with equation of state $p=-\epsilon$
 although many interesting proposals are discussed last time if $p\neq-\epsilon$. Besides, the equation of
 state of dark energy seems to be smaller than minus one as suggested by the cosmological data [11].
 Many theoretical problems arise when $w < -1$ [12], however, they can be overcome as
 V.Rubakov [13] has shown recently.

      Note also that other vacuum condensate (gravitational one) has fixed time in our
 Universe for $T \sim 10 ^{19}$ GeV [14].\\

 \vspace{0.1cm}

 References\\
 1.  News-release of STSci-2006-\textbf{52}.\\
 2.  V.Sahni,  Class. Quant.Grav. \textbf{19}, 3435 (2002).\\
 3.  T.Padmanabhan, Phys. Rep. \textbf{380}, 235 (2003).\\
 4.  L.Randall, The talk on KICP Inaugural Symp. memory of D.Schramm in December of 2005;\\ http://newviews.uchicago.edu/online-talks.html.\\
 5.  S.Hawking and T.Hertog, hep-th/0602091; Phys. Rev. D \textbf{73}, 123527 (2006).\\
 6.  V.Burdyuzha et al., Phys. Rev. D \textbf{55}, 7340R (1997).\\
 7.  E.Shuryak E. hep-ph/9503427; Phys. Rep. \textbf{264}, 357 (1996).\\
 8.  Ya.Zel'dovich, Pis'ma JETP \textbf{6}, 883 (1967).\\
 9.  N.Kardashev, Astron.Zh. \textbf{74}, 803 (1998).\\
 10. V.Burdyuzha, Proceedings of the Symp."Particles,\\Strings,Cosmology" (PASCOS-98)
       Ed. P.Nath, World Scientific, p.101 (1999).\\
 11. M.Tegmark, et al . Phys. Rev. D \textbf{74}, 123507 (2006).\\
 12. S.Carroll, M.Hoffman, M.Trodden, Phys. Rev. D \textbf{68}, 023509 (2003).\\
 13. V.Rubakov, hep-th/0604153.\\
 14. V.Burdyuzha and G.Vereshkov, Astrophys. Space Sci. \textbf{305}, 235 (2006).\\

\end{document}